\documentclass[11pt]{article}

\usepackage[margin=1in]{geometry}
\usepackage[T1]{fontenc}
\usepackage{lmodern}
\usepackage{microtype}

\usepackage{amsmath,amssymb,mathtools,bm}
\usepackage{graphicx}
\usepackage{booktabs}
\usepackage{multirow}
\usepackage{algorithm}
\usepackage{algpseudocode}

\usepackage[numbers,sort&compress]{natbib}
\usepackage{xcolor}
\usepackage[hidelinks]{hyperref}
\hypersetup{
  pdftitle={From a Static Multi-Level Small Semantic Codebook to a Dynamic Single-Level Large Semantic Codebook for Generative Recommendation},
  pdfauthor={Author Name}
}
\graphicspath{{figures/}{exp_sid_distribution/}}

\title{\textbf{From a Static Multi-Level Small Semantic Codebook to a Dynamic\\
Single-Level Large Semantic Codebook for Generative Recommendation}}
\author{Tianlu Xie$^{1}$, Xin Ku$^{1}$, Mingjie Sun$^{1}$,
Yunhao Sha$^{1}$, Lixiang Wang$^{1}$\\
Peng Wang$^{1}$, Yiyu Wang$^{1}$, Wenjin Wu$^{1}$,
Zhaojie Liu$^{1}$, Peng Jiang$^{1}$, Wenwu Ou$^{1}$\\
$^{1}$Kuaishou Technology, Beijing, China\\
\texttt{\{xietianlu, kuxin, sunmingjie, shayunhao,}\\
\texttt{wanglixiang03, wangpeng16, liubin21, wuwenjin,}\\
\texttt{zhaotianxing, jiangpeng, luocheng10\}@kuaishou.com}}
\date{}

\begin{document}

\maketitle

\begin{abstract}
Generative recommendation represents each item with a sequence of discrete Semantic IDs (SIDs) and predicts that sequence to retrieve the next item. Typical systems construct a multi-level semantic representation through residual quantization. In our industrial setting, the common three-level realization uses two semantic codes followed by a collaborative disambiguation code. The additional semantic level lengthens autoregressive decoding while creating a large hierarchical space that is sparsely occupied under each first-level code. Online systems also face \textbf{codebook drift}. As new items continually arrive and exposure distributions change, the partitions of a static codebook become increasingly misaligned with current traffic. We propose a \textbf{single-level large semantic codebook} that replaces the multi-level residual semantic representation with one semantic token and retains a separate collaborative disambiguation token to reduce item collisions. We further introduce an \textbf{exposure-aware dynamic codebook update mechanism} that combines temporal weight decay, exponential moving-average center updates, and an exposure-weighted penalty on SID changes. Finally, we develop an \textbf{offline evaluation framework} covering representation quality, semantic-code utilization, cluster load, full-SID collision, and temporal stability. Across the two public datasets, the two-level SID improves mean Recall@10 by \textbf{5.0\%--8.8\%} and mean NDCG@10 by \textbf{4.1\%--5.1\%} for OneRec-V1. The corresponding gains for OneRec-V2 are \textbf{7.1\%--8.7\%} and \textbf{3.8\%--8.5\%}. Under the fixed-date KuaiRec evaluation, dynamic updating further improves Recall@10 by \textbf{1.4\%} and \textbf{2.7\%}, and NDCG@10 by \textbf{7.0\%} and \textbf{2.7\%}, for OneRec-V1 and OneRec-V2, respectively. Across the decoder, LazyAR, and MTP serving architectures, the shorter SID reduces estimated autoregressive-decoding FLOPs by \textbf{47.93\%--48.70\%} and increases single-card QPS by \textbf{28.57\%--47.0\%}. In a five-day online A/B test serving 2.5\% of production traffic, the two-level SID improves the primary consumption metric by \textbf{0.792\%}.
\end{abstract}


\section{Introduction}
\label{sec:introduction}

Many large-scale recommender systems encode user queries and item candidates as continuous vectors and retrieve candidates through approximate nearest-neighbor search~\citep{rajput2023tiger}. Generative recommendation offers a different formulation: it represents each item as a sequence of discrete \textbf{Semantic IDs (SIDs)} and generates the SID sequence of the next item. TIGER established this paradigm by quantizing content embeddings into hierarchical semantic codewords and predicting them with an encoder--decoder Transformer~\citep{rajput2023tiger}. Subsequent studies have made item tokenization a central research problem, since the identifiers must preserve semantic structure, incorporate collaborative signals, and maintain balanced code usage~\citep{wang2024letter,chen2024selftoken,zheng2025mtgrec,zhai2025simcit,fu2026diger}.

Most SID-based generative recommenders construct hierarchical, multi-token item representations with \textbf{multi-level residual semantic quantization}~\citep{rajput2023tiger,wang2024letter,zheng2025mtgrec,hou2025rpg}. The first semantic code approximates the item embedding, and each subsequent semantic code quantizes the remaining residual to refine the representation. This design keeps the vocabulary at each decoding step manageable. However, an autoregressive recommender predicts the codes one step at a time and typically maintains multiple candidate paths with beam search. Every additional level therefore introduces another sequential prediction step and enlarges the search process~\citep{hou2025rpg}.

\begin{figure}[htbp]
  \centering
  \begin{minipage}[c]{0.61\linewidth}
    \centering
    \includegraphics[width=\linewidth]{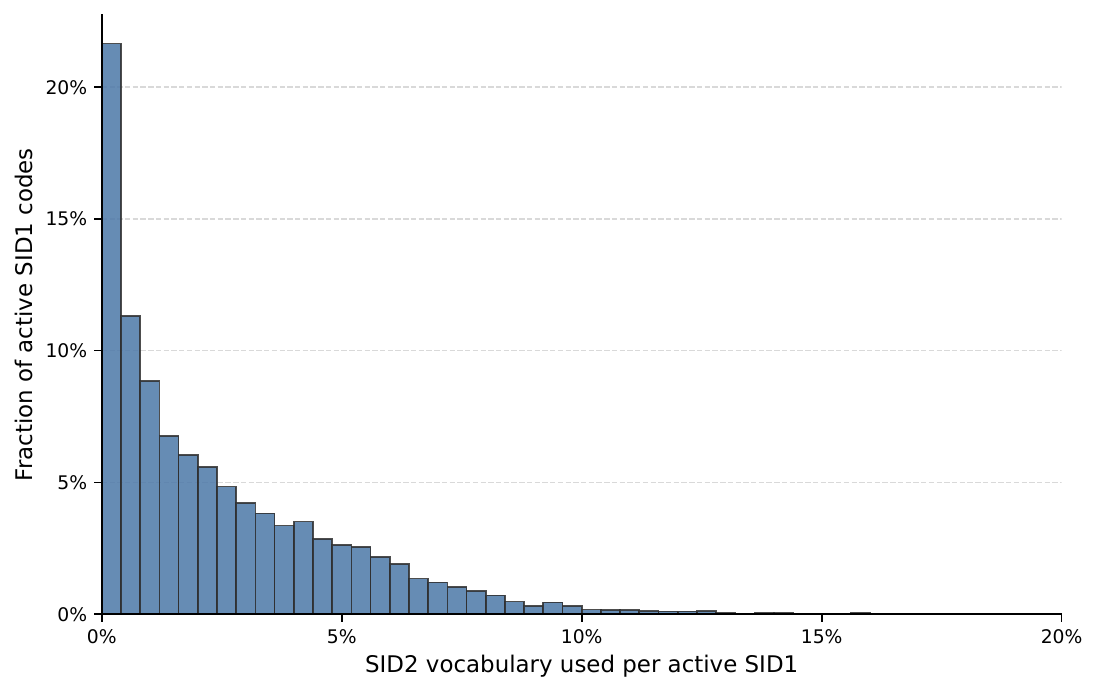}
  \end{minipage}\hfill
  \begin{minipage}[c]{0.35\linewidth}
    \centering
    \scriptsize
    \begin{tabular}{@{}lr@{}}
      \toprule
      Statistic & Value \\
      \midrule
      Samples & 1.50B \\
      Global SID1 utilization & 87.51\% \\
      Global SID2 utilization & 93.31\% \\
      \addlinespace
      Minimum & 0.02\% \\
      Maximum & 19.34\% \\
      Mean & 2.48\% \\
      99th percentile & 10.46\% \\
      95th percentile & 7.32\% \\
      90th percentile & 5.96\% \\
      75th percentile & 3.83\% \\
      Median & 1.68\% \\
      25th percentile & 0.50\% \\
      \bottomrule
    \end{tabular}
  \end{minipage}
  \caption{Global and conditional utilization of the two semantic levels in the deployed three-level SID. The histogram shows, for each active SID1, the fraction of the SID2 vocabulary observed under that code. The table reports global utilization and summarizes the conditional utilization distribution.}
  \label{fig:conditional_second_level_usage}
\end{figure}

In the industrial system considered in this work, a typical deployed \textbf{three-level SID} exposes a limitation of this standard design. SID1 and SID2 perform semantic residual quantization, whereas the collaborative SID3 only disambiguates items that remain indistinguishable under the semantic codes and thereby reduces full-SID collisions. An analysis of approximately $1.50$ billion industrial samples reveals a marked gap between the global coverage of semantic SID2 and its conditional use under SID1. Although 93.31\% of the SID2 vocabulary appears globally, each active SID1 is associated with only 2.48\% of that vocabulary on average. \textbf{SID2 therefore creates a large but conditionally sparse hierarchical space while still requiring an additional decoding step}. Figure~\ref{fig:conditional_second_level_usage} presents both the full distribution and its summary statistics.

Parallel SID generation has been explored as one way to reduce sequential dependencies, but independently predicted tokens are not guaranteed to compose a valid catalog identifier. RPG therefore uses graph-guided decoding to exclude invalid SIDs~\citep{hou2025rpg}, while LLaDA-Rec introduces adapted beam search for its diffusion decoder~\citep{shi2025lladarec}. Thus, parallel prediction still requires validity recovery or constrained search. We pursue a different direction: \textbf{reduce the number of semantic codes} that must be generated and avoid invalid cross-level combinations at the source.

Motivated by the large-vocabulary prediction capability of modern language models, we replace the multi-level residual semantic representation with a \textbf{single-level large semantic codebook}. A separate collaborative disambiguation code is retained only to reduce collisions. It is not part of semantic reconstruction. The resulting SID contains one semantic code and one disambiguation code, shortening the typical industrial sequence from three codes to two and reducing the number of autoregressive decoding steps.

Structural simplification alone does not address \textbf{codebook drift in a non-stationary online item corpus}. New items arrive continuously, item embeddings evolve, and exposure distributions shift with user interests and traffic allocation. A semantic codebook fitted to a fixed historical window can therefore become stale. Its centers and partitions no longer match the items that dominate current traffic. Recent studies have questioned static SID assignment, developed dynamic SIDs, or introduced adaptive item indexing for streaming recommendation~\citep{fu2026diger,shi2026ssrlive,yan2026merge}. We therefore introduce an \textbf{exposure-aware dynamic update mechanism} that maintains temporally decayed exposure weights, updates active centers with an exponential moving average, and penalizes exposure-weighted SID changes to improve the stability of high-exposure items.

In industrial systems, directly validating a new codebook through the complete recommendation pipeline is expensive. Each candidate codebook requires the item corpus to be retokenized and the downstream generative recommender to be retrained and evaluated. This long feedback loop makes online trial-and-error unsuitable for rapid codebook development. We therefore build a \textbf{codebook-level offline evaluation framework} that measures representation quality, semantic-code utilization, cluster load, full-SID collision, and temporal stability without requiring a full recommender training run for every candidate codebook. When a dataset provides meaningful content-type annotations, we additionally analyze the composition of those types within clusters. These diagnostics enable efficient codebook screening and comparison. Final recommendation quality is still evaluated by training generative recommenders on public datasets and industrial data.

Across the two public datasets, the two-level SID improves mean Recall@10 by \textbf{5.0\%--8.8\%} and mean NDCG@10 by \textbf{4.1\%--5.1\%} for OneRec-V1. The corresponding gains for OneRec-V2 are \textbf{7.1\%--8.7\%} and \textbf{3.8\%--8.5\%}. Under the fixed-date KuaiRec evaluation, dynamic updating further improves Recall@10 by \textbf{1.4\%} and \textbf{2.7\%}, and NDCG@10 by \textbf{7.0\%} and \textbf{2.7\%}, for OneRec-V1 and OneRec-V2, respectively. In industrial codebook evaluation, dynamic PV-S2 improves reconstruction similarity by \textbf{1.50\%} relative to static PV-S2. Across the decoder, LazyAR, and MTP serving architectures, estimated autoregressive-decoding FLOPs decrease by \textbf{47.93\%--48.70\%}, while single-card QPS increases by \textbf{28.57\%--47.0\%}. Finally, in a five-day online A/B test serving 2.5\% of production traffic, the two-level SID improves the primary consumption metric by \textbf{0.792\%}.

The main contributions of this work are as follows:
\begin{itemize}
  \item \textbf{Single-level large semantic codebook.} We consolidate multiple semantic quantization levels into one large semantic codebook and append a separate collaborative disambiguation code, shortening the complete SID and reducing autoregressive decoding steps.
  \item \textbf{Exposure-aware dynamic codebook.} We construct the dynamic codebook using temporally decayed exposure weights, exponential moving-average center updates, and an exposure-weighted SID-switching penalty, thereby balancing adaptation to changing traffic with assignment stability.
  \item \textbf{Offline codebook evaluation framework.} We develop multidimensional diagnostics that enable efficient codebook screening before the costly process of retokenizing items and retraining the downstream generative recommender.
\end{itemize}

\section{Related Work}
\label{sec:related_work}

\subsection{Semantic ID Tokenization and Codebook Design}
\label{sec:related_tokenization}

\textbf{Semantic ID construction.} Generative recommendation formulates next-item prediction as sequence generation over item identifiers, making identifier design central to both the model's output space and the structure shared across items. Atomic identifiers preserve item distinctiveness but expose no relations among items, whereas textual identifiers carry semantics but may not map cleanly and uniquely to catalog items~\citep{tay2022dsi,geng2022p5,chen2024selftoken}. Codebook-based Semantic IDs provide a compact, compositional alternative~\citep{wang2024letter}. VQ-VAE provides the underlying discretization mechanism by assigning encoder outputs to entries in a learned codebook~\citep{oord2017vqvae}. Residual quantization successively encodes the remaining approximation error, producing a multi-level code sequence without requiring one extremely large codebook~\citep{lee2022rqvae}. Adding residual levels can refine the representation, but it also lengthens the token sequence that the downstream generative model must predict. TIGER adopts this construction to quantize content embeddings into hierarchical SIDs for generative retrieval~\citep{rajput2023tiger}. Related work in generative retrieval learns discrete document identifiers end to end or organizes item identifiers in a balanced semantic tree~\citep{sun2023genret,si2024seater}.

\textbf{Recommendation-aligned tokenization.} In recommendation, SID quality depends on more than reconstruction fidelity. LC-Rec learns vector-quantized item indices with uniform semantic mapping and uses alignment tasks to integrate language and collaborative semantics into an LLM-based recommender~\citep{zheng2024lcrec}. LETTER regularizes an RQ-VAE tokenizer with objectives for hierarchical semantics, collaborative alignment, and code-assignment diversity~\citep{wang2024letter}. ETEGRec further couples the tokenizer and recommender through recommendation-oriented alignment objectives and alternating end-to-end optimization~\citep{liu2025etegrec}. Other studies refine initial item tokens using feedback from the generative model, augment pretraining with multiple RQ-VAE identifiers, or replace reconstruction with contrastive multimodal alignment~\citep{chen2024selftoken,zheng2025mtgrec,zhai2025simcit}. DIGER makes SID learning differentiable so that recommendation gradients can update the tokenizer, while using uncertainty-controlled exploration to mitigate codebook collapse~\citep{fu2026diger}. These methods establish item tokenization as a recommendation-aware learning problem rather than a purely semantic compression step.

\textbf{Codebook capacity.} Work on discrete generative modeling has also explored how to increase token capacity without relying exclusively on deeper code stacks. Finite scalar quantization constructs a large implicit codebook as a Cartesian product of scalar levels and avoids the learned-codebook collapse observed in conventional VQ~\citep{mentzer2024fsq}. Lookup-free quantization similarly supports large discrete vocabularies without an embedding lookup and has been used to build expressive tokenizers for autoregressive generation~\citep{yu2024magvit2}. Although these techniques use different quantizers and were developed outside recommendation, they show that discrete capacity can be increased along the vocabulary dimension rather than only through deeper code stacks. This width--depth trade-off is central to the SID architecture studied in this work.

\subsection{SID Generation and Constrained Decoding}
\label{sec:related_generation}

\textbf{Autoregressive SID generation.} Once items have been tokenized, a generative recommender predicts the code sequence of the target item from the user's interaction history. Hierarchical SIDs are commonly generated from left to right, with each prediction conditioned on the preceding codes~\citep{rajput2023tiger,si2024seater,zheng2025mtgrec}. Because only a small subset of all possible code sequences corresponds to catalog items, inference typically uses beam search together with catalog-derived constraints to retain valid candidates. This procedure couples the cost of inference to SID length: additional codes require additional model predictions and expand the sequence of search decisions~\citep{hou2025rpg}.

\textbf{Parallel and unordered generation.} RPG removes the fixed ordering among SID tokens, predicts a long SID in parallel, and uses graph-guided decoding to exclude invalid identifiers~\citep{hou2025rpg}. LLaDA-Rec replaces left-to-right decoding with discrete diffusion and introduces an adapted beam-search procedure for parallel SID generation~\citep{shi2025lladarec}. REG4Rec constructs multiple unordered semantic tokens with a mixture-of-experts-based parallel quantization codebook and uses them to support diverse reasoning paths, followed by consistency-oriented pruning~\citep{xing2025reg4rec}. These methods weaken token-level sequential dependence, but they still require additional mechanisms to ensure valid and reliable item predictions.

\subsection{Dynamic Semantic IDs and Streaming Item Indexing}
\label{sec:related_dynamic}

\textbf{Streaming clustering.} Conventional item tokenization assumes a fixed corpus and a stationary embedding distribution. Streaming settings violate both assumptions: new items arrive over time, while the relative prevalence and centroids of existing clusters may shift. Online clustering addresses continual arrival by assigning each observation and updating cluster centers incrementally~\citep{bhattacharjee2023onlinekmeans}. Forgetting factors further reduce the influence of older observations, allowing cluster statistics and quality estimates to reflect evolving structure~\citep{moshtaghi2019onlinevalidity}. For item indexing, however, adaptation must be balanced against assignment stability, because changing an item's SID also changes the discrete target consumed by downstream models.

\textbf{Adaptive recommendation indexing.} Recent recommender systems have begun to model this temporal dimension explicitly. SSRLive argues that static SIDs cannot represent rapidly changing live-room content and generates both static and dynamic SIDs within a unified generative--discriminative architecture~\citep{shi2026ssrlive}. MERGE targets skewed and non-stationary item distributions in large-scale streaming recommendation by constructing clusters adaptively, monitoring cluster occupancy, and forming hierarchical indices through fine-to-coarse merging~\citep{yan2026merge}.

\subsection{Evaluation of Semantic ID Codebooks}
\label{sec:related_evaluation}

\textbf{Codebook quality metrics.} Evaluation of vector quantizers commonly combines reconstruction fidelity with statistics of discrete-code usage. VQ-VAE evaluates the quality of the reconstructed input and the learned discrete latent representation, while subsequent work treats codebook collapse and inactive codes as separate diagnostic concerns~\citep{oord2017vqvae,mentzer2024fsq}. In generative recommendation, LETTER analyzes code-assignment and code-embedding distributions, relating imbalanced assignments to generation bias~\citep{wang2024letter}. DRQ provides a more explicit diagnostic framework based on expected codeword overlap and effective codebook capacity. Its industrial case study jointly examines latent geometry, robustness under perturbation, codebook utilization, reconstruction fidelity, and behavior-aware retrieval, illustrating that these dimensions need not improve together~\citep{wang2026drq}.

\textbf{Task-level and temporal evaluation.} Existing studies typically construct a codebook and assign SIDs before training a generative recommender, and then report downstream ranking metrics such as Recall and NDCG~\citep{rajput2023tiger,wang2024letter,chen2024selftoken,zheng2025mtgrec,liu2025etegrec}. This task-level protocol directly measures recommendation performance, but it offers limited evidence about which property of the SID construction method accounts for a performance difference. A separate line of streaming-clustering research develops online cluster-validity indices that update quality statistics as data arrive, enabling the monitoring of evolving cluster structures~\citep{moshtaghi2019onlinevalidity}. Existing evaluation practices therefore span codebook quality metrics, downstream recommendation accuracy, and temporal cluster validity, but an integrated evaluation protocol tailored to both the static quality and temporal behavior of recommendation SIDs remains underexplored.

\section{Method}
\label{sec:method}

\subsection{Problem Formulation and Overview}
\label{sec:method_overview}

Let $\mathcal{I}_t$ denote the item corpus available on day $t$. Each item $i\in\mathcal{I}_t$ is represented by a precomputed embedding $\mathbf{e}_i\in\mathbb{R}^{d}$, a stable item key $\mathbf{m}_i$, and an exposure count $p_i^{(t)}$. For semantic quantization, we use the $\ell_2$-normalized item embedding
\begin{equation}
  \mathbf{x}_i = \frac{\mathbf{e}_i}{\lVert \mathbf{e}_i\rVert_2}
  \in \mathbb{R}^{d}.
\end{equation}
The stable item key $\mathbf{m}_i$ is used only by the disambiguation rule described below.

Our framework produces a two-token SID
\begin{equation}
  \mathbf{s}_i^{(t)} = \bigl(s_{i,\mathrm{sem}}^{(t)}, s_{i,\mathrm{dis}}\bigr),
\end{equation}
where $s_{i,\mathrm{sem}}^{(t)}$ is selected from a large semantic codebook $\mathcal{C}^{(t)}=\{\mathbf{c}_k^{(t)}\}_{k=1}^{K_s}$ and $s_{i,\mathrm{dis}}$ is selected from a disambiguation vocabulary of size $K_d$. Throughout this paper, these two positions are referred to as \texttt{SID1} (semantic) and \texttt{SID2} (disambiguation), respectively. The method has three components. First, a single large semantic codebook replaces the two semantic residual-quantization levels in the original three-token SID while the existing disambiguation token is retained. Second, exposure-aware weights and temporally regularized center updates adapt the semantic codebook to a changing item distribution. Third, a codebook-level offline evaluation framework measures representation quality, semantic-code utilization, cluster load, full-SID collision, and temporal stability before downstream recommendation training.

\begin{figure*}[htbp]
  \centering
  \includegraphics[width=\textwidth]{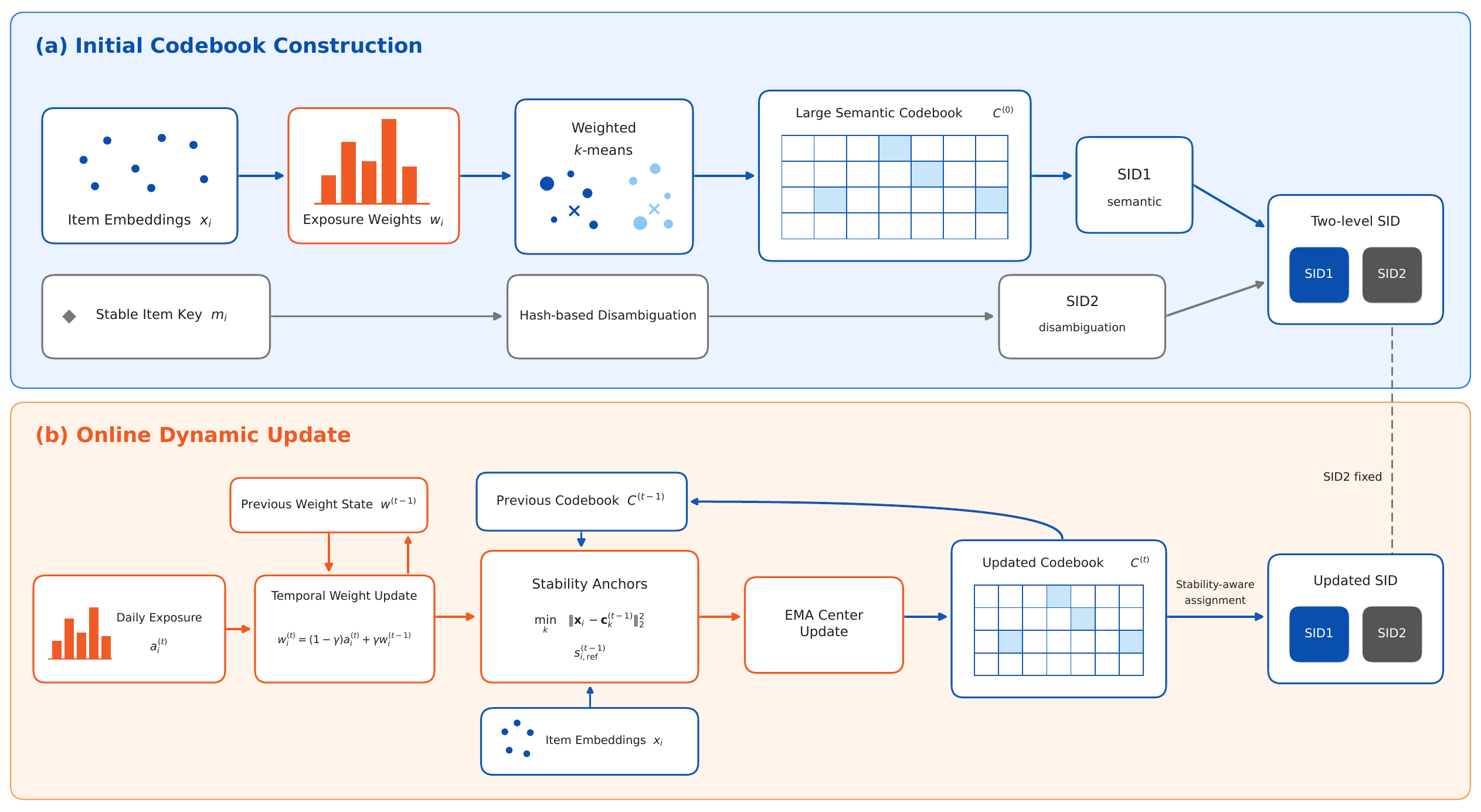}
  \caption{Overview of the proposed single-level large semantic codebook. Initial construction combines exposure-weighted semantic quantization (SID1) with deterministic collaborative disambiguation (SID2). Online updates maintain temporal exposure weights, penalize exposure-weighted SID1 switches, and update active centers while keeping SID2 unchanged.}
  \label{fig:method_pipeline}
\end{figure*}

Figure~\ref{fig:method_pipeline} summarizes initial construction and online updating. The upper panel learns the single large semantic codebook with exposure-weighted $k$-means and appends the deterministic disambiguation code. The lower panel updates only the semantic component. SID2 remains fixed across codebook versions.

\subsection{Single-Level Semantic Codebook with Disambiguation}
\label{sec:method_two_level}

\textbf{From two semantic tokens to one.} The original three-token architecture represents item $i$ as $(s_i^{(1)},s_i^{(2)},s_{i,\mathrm{dis}})$. Given semantic codebooks $\mathcal{C}^{(1)}$ and $\mathcal{C}^{(2)}$, residual quantization assigns the first code from $\mathbf{x}_i$ and the second from the remaining residual:
\begin{equation}
\begin{aligned}
  s_i^{(1)}
  &=\arg\min_k\left\lVert\mathbf{x}_i-\mathbf{c}_k^{(1)}\right\rVert_2^2,
  &\mathbf{r}_i^{(1)}
  &=\mathbf{x}_i-\mathbf{c}_{s_i^{(1)}}^{(1)},\\
  s_i^{(2)}
  &=\arg\min_k\left\lVert\mathbf{r}_i^{(1)}-\mathbf{c}_k^{(2)}\right\rVert_2^2,
  &\widehat{\mathbf{x}}_i^{\mathrm{RQ}}
  &=\mathbf{c}_{s_i^{(1)}}^{(1)}+\mathbf{c}_{s_i^{(2)}}^{(2)}.
\end{aligned}
\label{eq:original_rq}
\end{equation}
We instead concentrate the semantic capacity in one codebook. For the normalized item embedding $\mathbf{x}_i$ and semantic codebook $\mathcal{C}^{(t)}$, the assignment is
\begin{equation}
  s_{i,\mathrm{sem}}^{(t)}
  = \arg\min_{k\in\{1,\ldots,K_s\}}
  \left\lVert \mathbf{x}_i-\mathbf{c}_k^{(t)}\right\rVert_2^2,
  \qquad
  \widehat{\mathbf{x}}_i^{(t)}=\mathbf{c}_{s_{i,\mathrm{sem}}^{(t)}}^{(t)}.
  \label{eq:semantic_assignment}
\end{equation}
The large vocabulary increases the capacity of the single semantic token without reintroducing another autoregressive position.

\textbf{Disambiguation token.} Semantic quantization may assign multiple items to the same semantic code. We therefore retain the final disambiguation layer of the deployed system. Let $g$ be a fixed assignment rule over the stable item key:
\begin{equation}
  s_{i,\mathrm{dis}} = g(\mathbf{m}_i) \in \{1,\ldots,K_d\}.
  \label{eq:disambiguation}
\end{equation}
The rule is deterministic and held fixed across codebook updates. In the industrial implementation, $g(\mathbf{m}_i)$ is a stable hash over the item-associated key tuple $\mathbf{m}_i$. Consequently, an unchanged item receives the same disambiguation token even when its semantic assignment is updated. This token is used to separate items sharing a semantic code and does not participate in semantic reconstruction.

\textbf{Autoregressive generation.} For a user history $\mathcal{H}$, the proposed target factorizes as
\begin{equation}
  P(\mathbf{s}_i\mid\mathcal{H})
  =P(s_{i,\mathrm{sem}}\mid\mathcal{H})
   P(s_{i,\mathrm{dis}}\mid\mathcal{H},s_{i,\mathrm{sem}}).
\end{equation}
The original architecture requires an additional conditional factor for the second semantic token. Our design therefore reduces the number of autoregressive predictions from three to two while preserving the disambiguation stage.

\subsection{Exposure-Aware Codebook Learning}
\label{sec:method_weighted_learning}

Item frequency in industrial traffic is strongly skewed. An unweighted objective assigns the same importance to every item, regardless of its exposure frequency. The resulting partition therefore reflects the catalog distribution rather than the traffic distribution and may allocate insufficient representational capacity to frequently exposed items. To account for traffic while preventing head items from dominating the optimization, we compress the raw daily exposure count as follows:
\begin{equation}
  a_i^{(t)} = 1+\log_{10}\!\left(\max\{p_i^{(t)},1\}\right).
  \label{eq:daily_exposure}
\end{equation}
For a static training window, the resulting exposure weight is used in the weighted quantization objective
\begin{equation}
  \min_{\mathcal{C},\{s_i\}}
  \sum_i w_i
  \left\lVert\mathbf{x}_i-\mathbf{c}_{s_i}\right\rVert_2^2.
  \label{eq:weighted_quantization}
\end{equation}
The initial semantic codebook is trained with weighted $k$-means. During weighted $k$-means++ initialization, let $D_i^2=\min_{\mathbf{c}\in\mathcal{C}_{\mathrm{sel}}}\lVert\mathbf{x}_i-\mathbf{c}\rVert_2^2$ denote the squared distance to the nearest selected center. Let $I_{\mathrm{next}}$ denote the item index selected as the next center. It is sampled according to
\begin{equation}
  P(I_{\mathrm{next}}=i\mid\mathcal{C}_{\mathrm{sel}})
  =\frac{w_iD_i^2}{\sum_j w_jD_j^2}.
  \label{eq:weighted_kmeanspp}
\end{equation}

\subsection{Exposure-Aware Dynamic Codebook Update}
\label{sec:method_dynamic}

\textbf{Temporal exposure weights.} We maintain one stateful weight for every item observed recently. For an item present both in the historical state and on day $t$, its weight is updated as
\begin{equation}
  w_i^{(t)}=\gamma w_i^{(t-1)}+(1-\gamma)a_i^{(t)},
  \qquad 0<\gamma<1.
  \label{eq:weight_update_seen}
\end{equation}
A newly observed item is initialized with $w_i^{(t)}=a_i^{(t)}$. If a historical item is absent on day $t$, its weight decays as $w_i^{(t)}=\gamma w_i^{(t-1)}$. Items whose decayed weights fall below a small threshold $\varepsilon$ are removed from the maintained state. This mechanism retains recent exposure history while gradually forgetting items that no longer receive traffic.

\textbf{Stability anchors.} To discourage unnecessary SID changes, we use the assignments under the previous codebook $\mathcal{C}^{(t-1)}$ as stability anchors for the day-$t$ update. We encode each current embedding with the previous codebook:
\begin{equation}
  s_{i,\mathrm{ref}}^{(t-1)}=
  \arg\min_k\left\lVert\mathbf{x}_i-\mathbf{c}_k^{(t-1)}\right\rVert_2^2.
\end{equation}
These reference assignments define the previous partition used to estimate the current daily centers and provide the anchors for the switching penalty below.

\textbf{Center update and final assignment.} For center $k$, let $A_k^{(t)}=\{i:s_{i,\mathrm{ref}}^{(t-1)}=k\}$ be its reference-partition items on day $t$. We compute the exposure-weighted daily center
\begin{equation}
  \overline{\mathbf{c}}_k^{(t)}=
  \frac{\sum_{i\in A_k^{(t)}}w_i^{(t)}\mathbf{x}_i}
       {\sum_{i\in A_k^{(t)}}w_i^{(t)}}.
  \label{eq:daily_center}
\end{equation}
For a center receiving at least one assignment, the updated value is
\begin{equation}
  \mathbf{c}_k^{(t)}=
  (1-\alpha)\mathbf{c}_k^{(t-1)}
  +\alpha\overline{\mathbf{c}}_k^{(t)},
  \qquad 0<\alpha\leq 1.
  \label{eq:center_ema}
\end{equation}
An inactive center remains unchanged, $\mathbf{c}_k^{(t)}=\mathbf{c}_k^{(t-1)}$. We then compute the published semantic assignment against the updated centers with an exposure-weighted switching penalty:
\begin{equation}
  s_{i,\mathrm{sem}}^{(t)}=\arg\min_{k}
  \left[
    \left\lVert\mathbf{x}_i-\mathbf{c}_k^{(t)}\right\rVert_2^2
    +\lambda w_i^{(t)}
      \mathbb{I}\!\left(k\neq s_{i,\mathrm{ref}}^{(t-1)}\right)
  \right].
  \label{eq:stable_assignment}
\end{equation}
Here, $\lambda\geq 0$ controls the adaptation--stability trade-off. The penalty discourages SID changes for high-exposure items while still allowing reassignment when it provides a sufficiently large reduction in quantization distance. This stability is important in streaming training because abruptly reassigning a high-exposure item changes many discrete supervision targets at once. The resulting discontinuity can destabilize gradient updates and hinder model optimization. We update the codebook once per day and publish the new semantic assignments together with the unchanged disambiguation rule.

Algorithm~\ref{alg:dynamic_codebook_update} summarizes one online update interval.

\begin{algorithm}[htbp]
  \caption{Exposure-aware dynamic semantic-codebook update}
  \label{alg:dynamic_codebook_update}
  \begin{algorithmic}[1]
    \Require $\mathcal{I}_t$ with $\{\mathbf{x}_i,p_i^{(t)},\mathbf{m}_i\}$, $\mathcal{C}^{(t-1)}$, $\{w_i^{(t-1)}\}$, and $\gamma,\lambda,\alpha,\varepsilon$
    \Ensure $\mathcal{C}^{(t)}$ and $\{(\mathrm{SID1}_i,\mathrm{SID2}_i)\}$
    \ForAll{items in the union of the current corpus and weight state}
      \If{$i\in\mathcal{I}_t$}
        \State Compute $a_i^{(t)}$ and update or initialize $w_i^{(t)}$
      \Else
        \State Set $w_i^{(t)}\gets\gamma w_i^{(t-1)}$
      \EndIf
      \State Remove $i$ from the weight state if $w_i^{(t)}<\varepsilon$
    \EndFor
    \ForAll{$i\in\mathcal{I}_t$}
      \State Encode $\mathbf{x}_i$ with $\mathcal{C}^{(t-1)}$ to obtain $s_{i,\mathrm{ref}}^{(t-1)}$
    \EndFor
    \State Form $A_k^{(t)}=\{i\in\mathcal{I}_t:s_{i,\mathrm{ref}}^{(t-1)}=k\}$ for all $k$
    \For{$k=1$ to $K_s$}
      \If{$A_k^{(t)}\neq\varnothing$}
        \State Compute $\overline{\mathbf{c}}_k^{(t)}$ and update $\mathbf{c}_k^{(t)}$ using Eqs.~\eqref{eq:daily_center}--\eqref{eq:center_ema}
      \Else
        \State Set $\mathbf{c}_k^{(t)}\gets\mathbf{c}_k^{(t-1)}$
      \EndIf
    \EndFor
    \ForAll{$i\in\mathcal{I}_t$}
      \State Assign semantic SID1 $s_{i,\mathrm{sem}}^{(t)}$ using Eq.~\eqref{eq:stable_assignment}
    \EndFor
    \State Publish $\mathcal{C}^{(t)}$ and $(\text{SID1},\text{SID2})=(s_{i,\mathrm{sem}}^{(t)},g(\mathbf{m}_i))$
  \end{algorithmic}
\end{algorithm}

\subsection{Offline Codebook Evaluation Framework}
\label{sec:method_evaluation}

Training a downstream generative recommender remains the definitive evaluation of recommendation quality, but it is unnecessarily expensive for every intermediate codebook candidate. We therefore evaluate codebooks first using a set of offline metrics. Let $\mathcal{E}_t$ be the evaluation items for day $t$ and $N_t=|\mathcal{E}_t|$.

\textbf{Representation quality.} We measure the cosine similarity between each normalized item embedding and its reconstruction:
\begin{equation}
  R_{\mathrm{rec}}=
  \frac{1}{N_t}\sum_{i\in\mathcal{E}_t}
  \cos\!\left(\mathbf{x}_i,\widehat{\mathbf{x}}_i\right).
  \label{eq:reconstruction_metric}
\end{equation}
Here, $\widehat{\mathbf{x}}_i$ is the selected semantic center. For a residual-quantization baseline, it is the sum of the selected semantic centers.
In addition to the mean, we report the 10th percentile (P10) and median of item-level reconstruction similarity to characterize its lower tail and central tendency.

\textbf{Utilization and cluster load.} For semantic level $\ell$ with vocabulary size $K_\ell$, let $q_i^{(\ell)}$ be the code assigned to item $i$. Its utilization is
\begin{equation}
  U_\ell=\frac{|\{q_i^{(\ell)}:i\in\mathcal{E}_t\}|}{K_\ell}.
\end{equation}
For load statistics, let $\sigma_i$ denote the complete semantic group of item $i$, namely one semantic code for the proposed structure and the tuple of semantic codes for a residual-quantization baseline. For each occupied group $k$, its item load is
\begin{equation}
  n_k=\sum_{i\in\mathcal{E}_t}\mathbb{I}(\sigma_i=k).
  \label{eq:semantic_group_load}
\end{equation}
The 95th percentile and maximum of $\{n_k\}$ identify crowded semantic groups, while the coefficient of variation (CV) measures overall load imbalance, with a smaller CV indicating a more even distribution. When a dataset provides meaningful content-type labels, we also report the type composition within each cluster. This analysis is omitted for datasets without such annotations.

\textbf{Full-SID collision.} Let $\mathcal{G}_z$ be the set of distinct items assigned to complete SID $z=(s_{\mathrm{sem}},s_{\mathrm{dis}})$, and let $\mathcal{Z}$ be the occupied SIDs. We report both the fraction of occupied SIDs that collide and the fraction of items involved in a collision:
\begin{equation}
  R_{\mathrm{SID}}=
  \frac{\sum_{z\in\mathcal{Z}}\mathbb{I}(|\mathcal{G}_z|>1)}{|\mathcal{Z}|},
  \qquad
  R_{\mathrm{item}}=
  \frac{\sum_{z\in\mathcal{Z}}|\mathcal{G}_z|\mathbb{I}(|\mathcal{G}_z|>1)}{N_t}.
  \label{eq:collision_metrics}
\end{equation}

\textbf{Temporal stability.} To isolate codebook change from feature change, the old and new codebooks encode the same target-day embeddings. Denote the resulting semantic codes by $q_i^{\mathrm{old}}$ and $q_i^{\mathrm{new}}$. The item-level and exposure-weighted change rates are
\begin{equation}
  R_{\mathrm{chg}}=
  \frac{1}{N_t}\sum_{i\in\mathcal{E}_t}
  \mathbb{I}(q_i^{\mathrm{old}}\neq q_i^{\mathrm{new}}),
  \qquad
  R_{\mathrm{chg}}^{\mathrm{PV}}=
  \frac{\sum_{i\in\mathcal{E}_t}p_i^{(t)}\mathbb{I}(q_i^{\mathrm{old}}\neq q_i^{\mathrm{new}})}
       {\sum_{i\in\mathcal{E}_t}p_i^{(t)}}.
  \label{eq:stability_metrics}
\end{equation}
These offline measurements are used to screen codebook candidates. Downstream recommendation experiments are still required to establish their task-level effectiveness.

\section{Experiments}
\label{sec:experiments}

\subsection{Experimental Setup}
\label{sec:exp_setup}

\textbf{Datasets.} We use Amazon Reviews 2014 (Beauty)~\citep{mcauley2015image} and the KuaiRec 2.0 Big Matrix~\citep{gao2022kuairec}. Their statistics are summarized in Table~\ref{tab:dataset_statistics}. Amazon Beauty is used for the controlled static S3--S2 comparison. KuaiRec provides dense timestamped viewing logs and supports both a static comparison with exposure-weighted S2 (PV-S2) and a separate fixed-date comparison between static and dynamic PV-S2.

\begin{table}[htbp]
  \centering
  \small
  \caption{Statistics of the public datasets used in our experiments.}
  \label{tab:dataset_statistics}
  \begin{tabular}{@{}lrrrl@{}}
    \toprule
    Dataset & Users & Items & Interactions & Temporal coverage \\
    \midrule
    Amazon Beauty (5-core) & 22,363 & 12,101 & 198,502 & May 1996--July 2014 \\
    KuaiRec 2.0 Big Matrix & 7,176 & 10,728 & 12,530,806 & 28 observed days \\
    \bottomrule
  \end{tabular}
\end{table}

\textbf{Data preprocessing and evaluation.} Interactions are ordered chronologically for each user, and KuaiRec retains viewing events with \texttt{watch\_ratio} $\geq 1.0$ from users with at least five retained events. Amazon and the first KuaiRec comparison use leave-two-out evaluation. The last interaction is used for testing, the penultimate interaction for validation, and earlier interactions for training. All KuaiRec viewing events define the PV weights used by the leave-two-out PV-S2 baseline. Because users' final interactions occur on different calendar dates, leave-two-out does not provide a common temporal boundary between codebook updating and evaluation. The second KuaiRec comparison therefore uses the disjoint fixed-date evaluation described in Section~\ref{sec:exp_kuairec_temporal}. The two sets of results are reported separately.

\textbf{Codebook sizes.} On both Amazon Beauty and KuaiRec, S3 uses level sizes $[256,256,256]$, whereas S2, PV-S2, and Dynamic PV-S2 use $[1024,512]$. The final level is a balanced disambiguation code. PV weighting and dynamic updates change how the semantic codebook is fitted, not its dimensions.

\textbf{Recommendation models.} We prioritize the OneRec-V1 and OneRec-V2 public-data adaptations~\citep{deng2025onerec,zhou2025onerecv2}, and additionally report TIGER~\citep{rajput2023tiger}, RPG~\citep{hou2025rpg}, SEATER~\citep{si2024seater}, and COBRA~\citep{yang2025cobra}. These models cover autoregressive, parallel, tree-constrained, and sparse--dense cascaded generative recommendation. OneRec is evaluated at nominal scales from 0.1B to 0.7B. The exact configurations are listed in Appendix~\ref{app:onerec_configs}. The public-data adaptations exclude production-only features and preference-alignment stages. Within each paired comparison, we keep the data split, recommender architecture, optimization settings, and seed fixed and change only the SID artifact.

\textbf{Metrics and repeated runs.} Recommendation quality is measured by Recall@$K$ and NDCG@$K$ for $K\in\{5,10\}$. Main public-data results use the same three seeds for every paired codebook variant and report the mean and sample standard deviation. Codebook evaluation follows Section~\ref{sec:method_evaluation}.

\subsection{Static Two-Level versus Three-Level SIDs on Amazon Beauty}
\label{sec:exp_amazon}

The complete OneRec-V1/V2 sweeps are summarized in Figure~\ref{fig:amazon_onerec_scaling}. Their numerical results are provided in Appendix~\ref{app:onerec_amazon_results}. A line plot is used because model scale is ordered. Shaded regions show the sample standard deviation across seeds. Table~\ref{tab:amazon_static_results} reports the remaining models retained in the current comparison.

\begin{figure}[htbp]
  \centering
  \includegraphics[width=\linewidth]{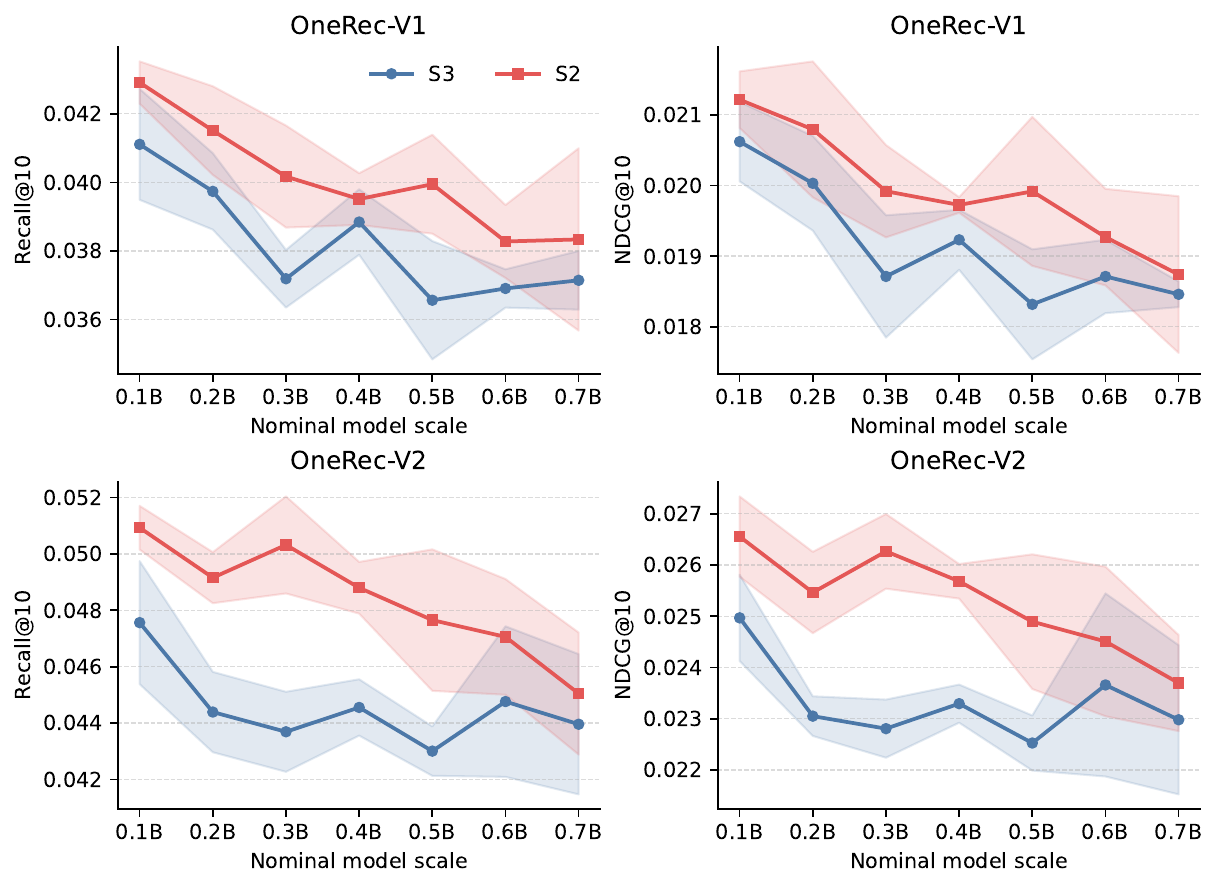}
  \caption{OneRec-V1/V2 scaling results on Amazon Reviews 2014 (Beauty). Curves and shaded regions denote the mean and sample standard deviation over seeds $\{7,42,2024\}$, respectively.}
  \label{fig:amazon_onerec_scaling}
\end{figure}

\begin{table}[htbp]
  \centering
  \small
  \caption{Recommendation performance on Amazon Reviews 2014 (Beauty). Results are mean $\pm$ sample standard deviation over seeds $\{7,42,2024\}$.}
  \label{tab:amazon_static_results}
  \resizebox{\linewidth}{!}{%
  \begin{tabular}{@{}llcccc@{}}
    \toprule
    Model & SID & Recall@5 & Recall@10 & NDCG@5 & NDCG@10 \\
    \midrule
    TIGER & S3 & $0.0369\pm0.0078$ & $0.0520\pm0.0136$ & $0.0253\pm0.0051$ & $0.0302\pm0.0070$ \\
    TIGER & S2 & $0.0328\pm0.0009$ & $0.0463\pm0.0008$ & $0.0230\pm0.0010$ & $0.0273\pm0.0010$ \\
    \addlinespace
    SEATER & S3 & $0.0380\pm0.0014$ & $0.0562\pm0.0014$ & $0.0256\pm0.0009$ & $0.0314\pm0.0010$ \\
    SEATER & S2 & $0.0385\pm0.0002$ & $0.0584\pm0.0008$ & $0.0258\pm0.0005$ & $0.0323\pm0.0006$ \\
    \addlinespace
    RPG & S3 & $0.0402\pm0.0004$ & $0.0569\pm0.0008$ & $0.0284\pm0.0003$ & $0.0337\pm0.0005$ \\
    RPG & S2 & $0.0441\pm0.0009$ & $0.0646\pm0.0006$ & $0.0310\pm0.0007$ & $0.0375\pm0.0006$ \\
    \addlinespace
    COBRA & S3 & $0.0062\pm0.0003$ & $0.0102\pm0.0001$ & $0.0040\pm0.0001$ & $0.0053\pm0.0000$ \\
    COBRA & S2 & $0.0103\pm0.0013$ & $0.0159\pm0.0014$ & $0.0069\pm0.0010$ & $0.0086\pm0.0010$ \\
    \bottomrule
  \end{tabular}}
\end{table}

Across all seven model scales, S2 improves mean Recall@10 over S3 by 5.0\% for OneRec-V1 and 8.7\% for OneRec-V2. The corresponding mean NDCG@10 gains are 4.1\% and 8.5\%. S2 is better at every evaluated scale for both OneRec variants on Recall@10. Among the other recommenders, S2 improves Recall@10 for SEATER, RPG, and COBRA, while TIGER performs better with S3. These results indicate that removing one semantic level generally preserves or improves recommendation quality, although the outcome remains model dependent.

\subsection{Static and Dynamic Codebooks on KuaiRec}
\label{sec:exp_kuairec}

\subsubsection{Static Comparison under Leave-Two-Out}

\paragraph{Offline codebook quality.}

Table~\ref{tab:kuairec_codebook_quality} reports offline codebook quality for the leave-two-out comparison. For S3, reconstruction uses the sum of the two selected semantic centers, whereas S2 and PV-S2 use one semantic center. S2 achieves the highest mean reconstruction cosine similarity, followed by S3 and PV-S2, and every semantic level retains at least 94.92\% utilization. Among the single-level variants, PV-S2 reduces the maximum group load from 376 to 248 and the load CV from 1.9938 to 1.6390.

\begin{table}[htbp]
  \centering
  \small
  \caption{Offline codebook quality on KuaiRec.}
  \label{tab:kuairec_codebook_quality}
  \resizebox{\linewidth}{!}{%
  \begin{tabular}{@{}lccc@{}}
    \toprule
    Metric & S3 & S2 & PV-S2 \\
    \midrule
    Mean reconstruction cosine similarity & 0.9958 & 0.9962 & 0.9939 \\
    Reconstruction P10 / median & 0.9879 / 0.9991 & 0.9879 / 0.9996 & 0.9823 / 0.9995 \\
    Semantic-code utilization & \shortstack{L1: 100.00\%\\L2: 100.00\%} & L1: 97.56\% & L1: 94.92\% \\
    Semantic-group load P95 / maximum / CV & 9.0 / 256 / 3.1685 & 26.1 / 376 / 1.9938 & 34.4 / 248 / 1.6390 \\
    \bottomrule
  \end{tabular}}
\end{table}

\paragraph{Recommendation performance.}

The complete OneRec-V1/V2 sweeps over the three static SID variants are shown in Figure~\ref{fig:kuairec_onerec_scaling}. Their numerical values are provided in Appendix~\ref{app:onerec_kuairec_results}. As in the Amazon comparison, curves report the three-seed mean and shaded regions show the sample standard deviation. Table~\ref{tab:kuairec_other_recommenders} reports the remaining models.

\begin{figure}[htbp]
  \centering
  \includegraphics[width=\linewidth]{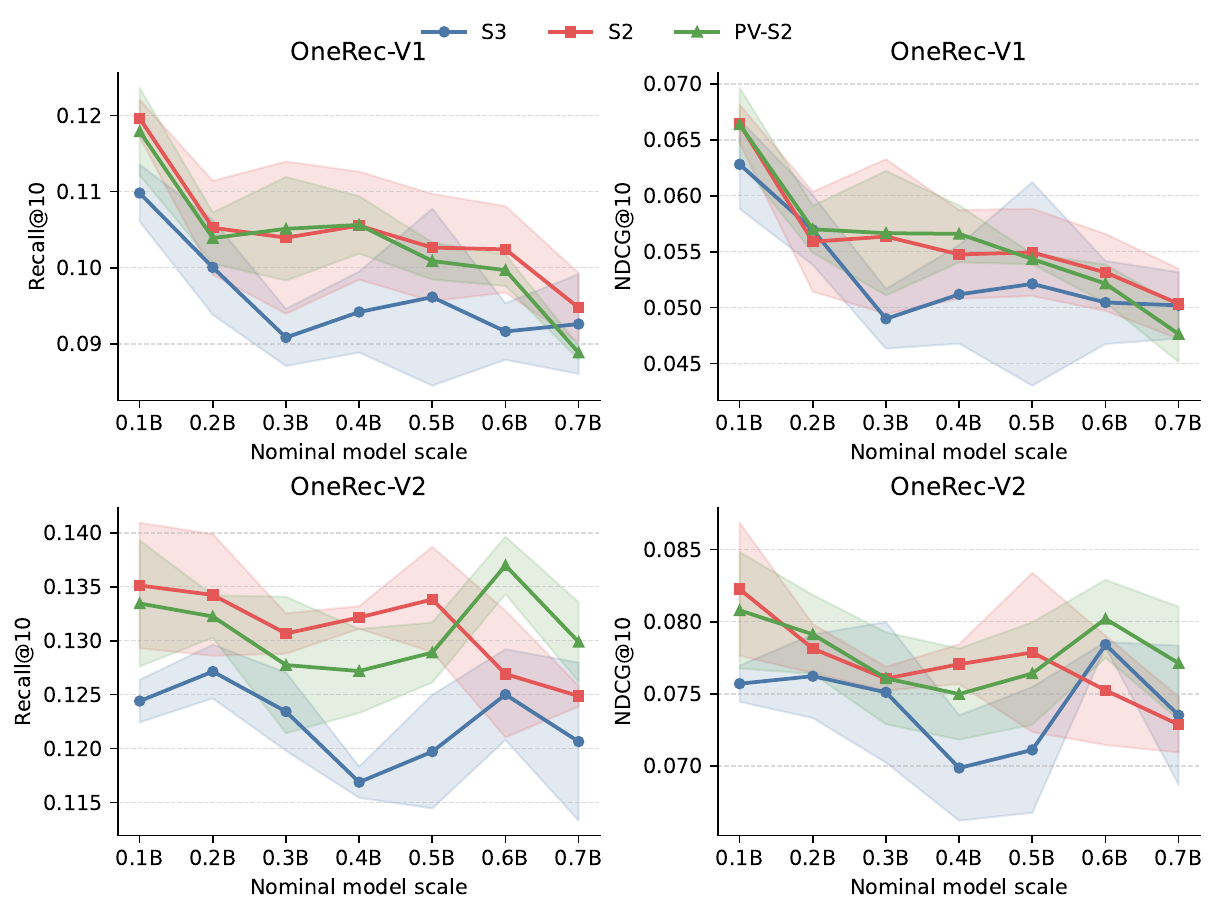}
  \caption{OneRec-V1/V2 scaling results on KuaiRec under S3, S2, and PV-S2. Curves and shaded regions denote the mean and sample standard deviation over seeds $\{7,42,2024\}$, respectively.}
  \label{fig:kuairec_onerec_scaling}
\end{figure}

\begin{table}[htbp]
  \centering
  \small
  \caption{Recommendation performance of the completed SID comparisons on KuaiRec. Results are mean $\pm$ sample standard deviation over seeds $\{7,42,2024\}$.}
  \label{tab:kuairec_other_recommenders}
  \resizebox{\linewidth}{!}{%
  \begin{tabular}{@{}llcccc@{}}
    \toprule
    Model & SID & Recall@5 & Recall@10 & NDCG@5 & NDCG@10 \\
    \midrule
    TIGER & S3 & $0.1057\pm0.0032$ & $0.1429\pm0.0019$ & $0.0740\pm0.0010$ & $0.0860\pm0.0006$ \\
    TIGER & S2 & $0.1094\pm0.0022$ & $0.1483\pm0.0012$ & $0.0761\pm0.0027$ & $0.0886\pm0.0023$ \\
    TIGER & PV-S2 & $0.1087\pm0.0028$ & $0.1492\pm0.0024$ & $0.0742\pm0.0030$ & $0.0873\pm0.0028$ \\
    \addlinespace
    SEATER & S3 & $0.1032\pm0.0019$ & $0.1374\pm0.0007$ & $0.0697\pm0.0028$ & $0.0808\pm0.0022$ \\
    SEATER & S2 & $0.1077\pm0.0018$ & $0.1473\pm0.0040$ & $0.0729\pm0.0019$ & $0.0858\pm0.0002$ \\
    SEATER & PV-S2 & $0.1069\pm0.0022$ & $0.1478\pm0.0028$ & $0.0717\pm0.0022$ & $0.0849\pm0.0024$ \\
    \addlinespace
    RPG & S3 & $0.0060\pm0.0004$ & $0.0082\pm0.0010$ & $0.0046\pm0.0003$ & $0.0053\pm0.0003$ \\
    RPG & S2 & $0.0060\pm0.0004$ & $0.0110\pm0.0020$ & $0.0037\pm0.0002$ & $0.0053\pm0.0006$ \\
    RPG & PV-S2 & $0.0109\pm0.0014$ & $0.0172\pm0.0017$ & $0.0066\pm0.0005$ & $0.0087\pm0.0008$ \\
    \addlinespace
    COBRA & S3 & $0.0998\pm0.0008$ & $0.1367\pm0.0011$ & $0.0693\pm0.0008$ & $0.0812\pm0.0005$ \\
    COBRA & S2 & $0.1088\pm0.0010$ & $0.1484\pm0.0034$ & $0.0754\pm0.0003$ & $0.0881\pm0.0009$ \\
    COBRA & PV-S2 & $0.1075\pm0.0037$ & $0.1486\pm0.0030$ & $0.0749\pm0.0016$ & $0.0882\pm0.0014$ \\
    \bottomrule
  \end{tabular}}
\end{table}

For OneRec-V1, averaging over seven model scales, S2 and PV-S2 improve Recall@10 over S3 by 8.8\% and 7.0\%, respectively, with corresponding NDCG@10 gains of 5.1\% and 4.8\%. For OneRec-V2, S2 and PV-S2 improve Recall@10 by 7.1\% and 6.9\%, and NDCG@10 by 3.8\% and 4.8\%, respectively. Both two-level variants also improve Recall@10 for every non-OneRec model in Table~\ref{tab:kuairec_other_recommenders}. Neither S2 nor PV-S2 dominates uniformly: exposure weighting produces the best Recall@10 for TIGER, SEATER, RPG, and COBRA, but the preferred variant varies across OneRec scales and ranking metrics.

\subsubsection{Fixed-Date Static--Dynamic Comparison}
\label{sec:exp_kuairec_temporal}

\begin{figure}[htbp]
  \centering
  \includegraphics[width=\linewidth]{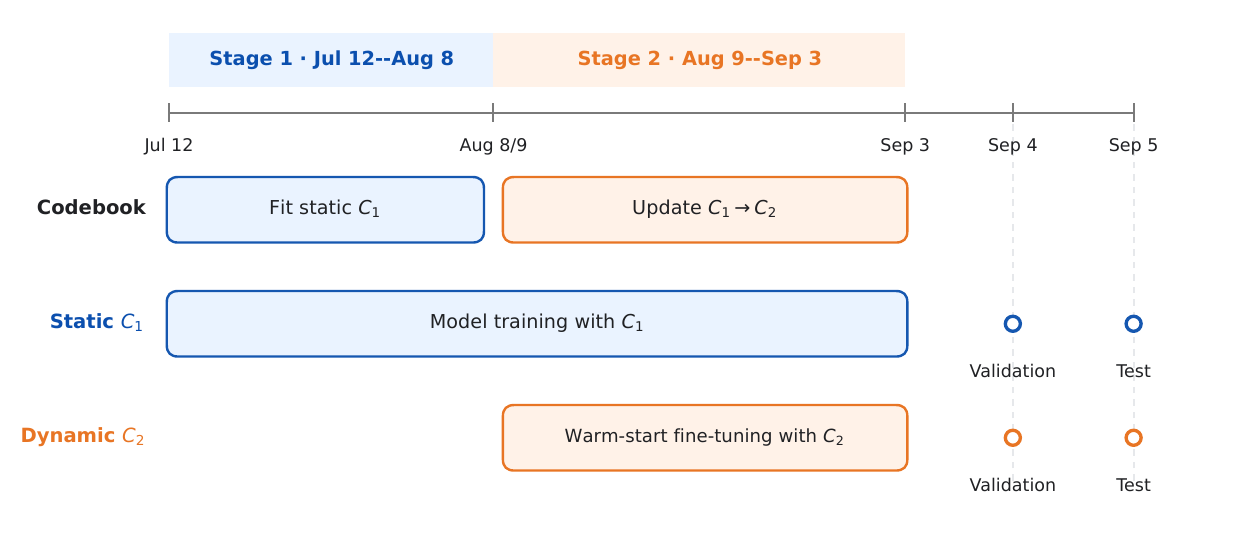}
  \caption{Fixed-date evaluation setup for the static--dynamic comparison on KuaiRec. Bars indicate the data windows used for codebook fitting or updating and recommender training. Both arms use the first positive interaction per user on September~4 for validation and on September~5 for testing.}
  \label{fig:kuairec_temporal_protocol}
\end{figure}

Figure~\ref{fig:kuairec_temporal_protocol} summarizes the fixed-date evaluation setup. The static arm trains with $\mathcal{C}_1$ on targets from July~12 through September~3. The dynamic arm updates $\mathcal{C}_1$ to $\mathcal{C}_2$ using traffic from August~9 through September~3, then fine-tunes from the matched static checkpoint on $\mathcal{C}_2$-encoded targets over the same period. Both arms validate on September~4 and test on September~5. The validation and test days are excluded from codebook fitting.

\begin{figure}[htbp]
  \centering
  \includegraphics[width=\linewidth]{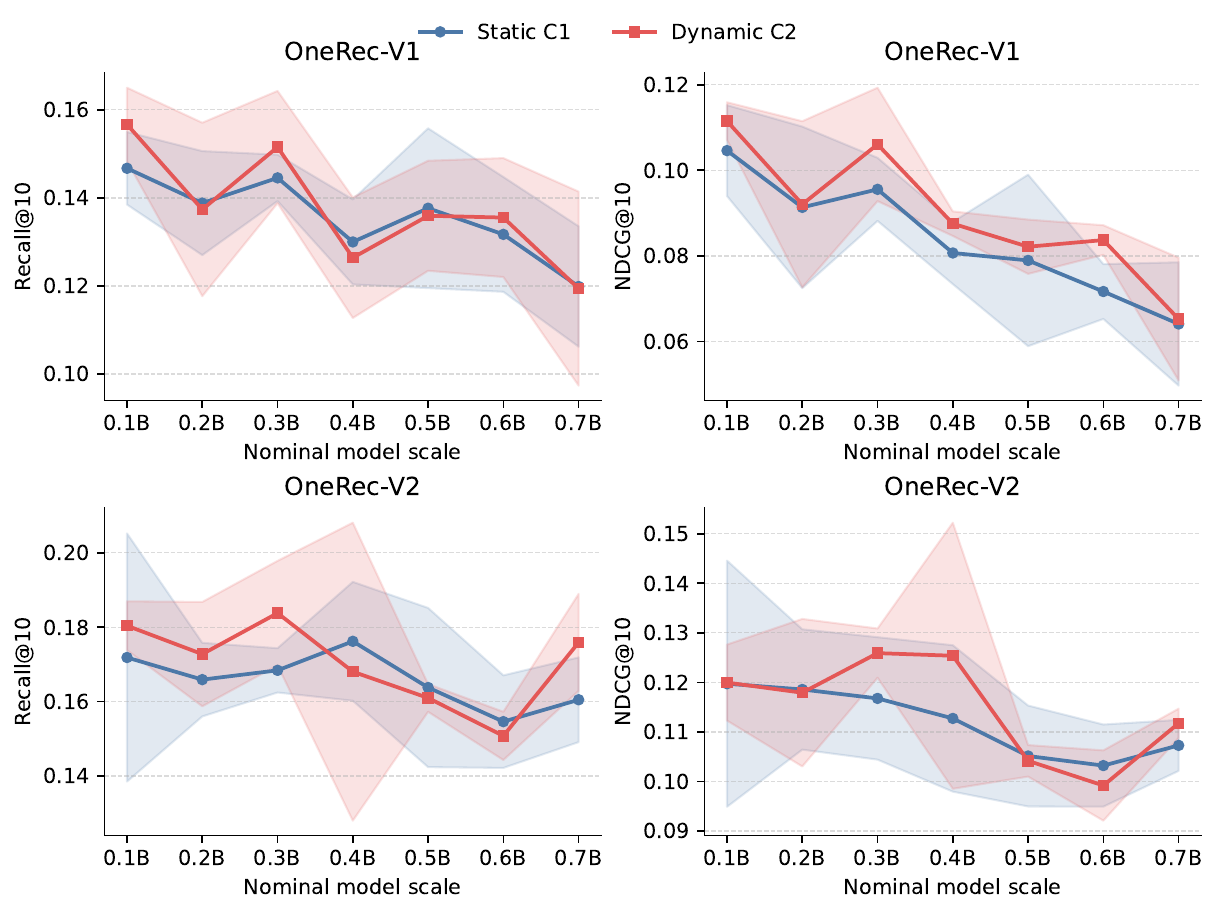}
  \caption{OneRec-V1/V2 results under the fixed-date KuaiRec comparison. Curves and shaded regions denote the mean and sample standard deviation over seeds $\{7,42,2024\}$, respectively. Static C1 uses the first-stage PV-S2 codebook, while Dynamic C2 uses the updated codebook and matched warm-start fine-tuning.}
  \label{fig:kuairec_onerec_temporal_scaling}
\end{figure}

Figure~\ref{fig:kuairec_onerec_temporal_scaling} compares the matched static and dynamic arms across OneRec scales. The exact numerical results are provided in Appendix~\ref{app:onerec_kuairec_temporal_results}. Table~\ref{tab:kuairec_temporal_other_recommenders} reports the completed three-seed comparisons for the remaining recommenders.

\begin{table}[htbp]
  \centering
  \small
  \caption{Fixed-date static--dynamic recommendation performance on KuaiRec. Results are mean $\pm$ sample standard deviation over seeds $\{7,42,2024\}$.}
  \label{tab:kuairec_temporal_other_recommenders}
  \resizebox{\linewidth}{!}{%
  \begin{tabular}{@{}llcccc@{}}
    \toprule
    Model & Codebook & Recall@5 & Recall@10 & NDCG@5 & NDCG@10 \\
    \midrule
    TIGER & Static C1 & $0.0075\pm0.0025$ & $0.0092\pm0.0011$ & $0.0048\pm0.0009$ & $0.0053\pm0.0004$ \\
    TIGER & Dynamic C2 & $0.0496\pm0.0775$ & $0.0532\pm0.0812$ & $0.0372\pm0.0584$ & $0.0383\pm0.0596$ \\
    \addlinespace
    SEATER & Static C1 & $0.1484\pm0.0014$ & $0.1788\pm0.0033$ & $0.1161\pm0.0034$ & $0.1260\pm0.0031$ \\
    SEATER & Dynamic C2 & $0.1570\pm0.0106$ & $0.1890\pm0.0090$ & $0.1215\pm0.0066$ & $0.1331\pm0.0068$ \\
    \addlinespace
    RPG & Static C1 & $0.0356\pm0.0238$ & $0.0669\pm0.0312$ & $0.0206\pm0.0151$ & $0.0307\pm0.0164$ \\
    RPG & Dynamic C2 & $0.0465\pm0.0136$ & $0.0688\pm0.0105$ & $0.0335\pm0.0073$ & $0.0392\pm0.0073$ \\
    \addlinespace
    COBRA & Static C1 & $0.0704\pm0.0086$ & $0.1116\pm0.0103$ & $0.0402\pm0.0072$ & $0.0551\pm0.0050$ \\
    COBRA & Dynamic C2 & $0.0892\pm0.0061$ & $0.1242\pm0.0037$ & $0.0558\pm0.0030$ & $0.0651\pm0.0027$ \\
    \bottomrule
  \end{tabular}}
\end{table}

Averaged across model scales, Dynamic C2 improves OneRec-V1 Recall@10 by 1.4\% and NDCG@10 by 7.0\% relative to Static C1. For OneRec-V2, the corresponding gains are 2.7\% and 2.7\%. Dynamic C2 also improves mean Recall@10 by 5.7\% for SEATER and 2.8\% for RPG. Although the gains are not uniform at every OneRec scale, the aggregate results are positive for both OneRec variants, SEATER, and RPG.

Relative to C1, C2 changes SID1 for 46 of 10,728 items, corresponding to an item-level change rate of 0.4288\%. The change rate is 0.6418\% among items active during the second stage and 0.4383\% when weighted by second-stage PV. The mean and maximum center displacements are 0.004054 and 0.050005, respectively. These measurements indicate that the update adapts the semantic partition while keeping SID1 assignments largely stable.

\subsection{Industrial Codebook Evaluation}
\label{sec:exp_industrial}

We evaluate the industrial codebooks on seven consecutive daily snapshots. For S2, reconstruction uses its single semantic center. For S3, we report both its first semantic level and the cumulative reconstruction from its two residual semantic levels. Table~\ref{tab:industrial_reconstruction_summary} shows the first and last daily means. Static S2 slightly exceeds the cumulative two-level semantic reconstruction of S3 on both endpoints. Static PV-S2 provides a further gain. Dynamic PV-S2 shares the same initialization and Day~1 value as static PV-S2, but reaches 0.826723 on Day~7 instead of declining with the static codebook.

\begin{table}[htbp]
  \centering
  \small
  \caption{Mean reconstruction similarity on the first and seventh industrial evaluation days.}
  \label{tab:industrial_reconstruction_summary}
  \begin{tabular}{@{}lrrr@{}}
    \toprule
    Codebook representation & Day 1 & Day 7 & Change \\
    \midrule
    Static S3, first semantic level & 0.758357 & 0.754451 & $-0.003906$ \\
    Static S3, two semantic levels & 0.802241 & 0.798868 & $-0.003373$ \\
    Static S2 & 0.805676 & 0.800899 & $-0.004777$ \\
    Static PV-S2 & 0.821336 & 0.814508 & $-0.006828$ \\
    Dynamic PV-S2 & 0.821336 & 0.826723 & $+0.005387$ \\
    \bottomrule
  \end{tabular}
\end{table}

\subsection{Industrial Online and Serving Evaluation}
\label{sec:exp_online_serving}

We deployed the two-level SID system in a five-day online A/B test serving 2.5\% of production traffic against the deployed three-level SID baseline. The primary consumption metric increased by 0.792\%. Across the decoder, LazyAR, and MTP serving architectures, estimated autoregressive-decoding FLOPs decreased by 47.93\%--48.70\%. Single-card QPS increased by 47.0\% for decoder, 28.57\% for LazyAR, and 33.3\% for MTP. The estimate includes the one-time cross-attention key/value projections, beam-expanded decoder computation, and vocabulary projection. Appendix~\ref{app:serving_cost} provides the symbolic derivation and separately characterizes KV-cache access. Together, the reduced decoding cost and measured throughput gains demonstrate the serving benefit of the shorter SID across decoding architectures.

\section{Conclusion}
\label{sec:conclusion}

This paper revisits Semantic ID codebook design for large-scale generative recommendation. Motivated by the sparse conditional use of the second semantic level in a deployed three-level SID, we replace the multi-level residual semantic representation with one large semantic codebook and retain a separate collaborative code only for disambiguation. The resulting semantic-plus-disambiguation SID concentrates semantic capacity in a single token and reduces the number of autoregressive predictions from three to two. To accommodate an evolving online item corpus, we further develop an exposure-aware dynamic codebook that combines temporally decayed exposure weights, stability-aware assignment, and moving-average center updates. Finally, we introduce an offline evaluation framework covering reconstruction quality, code utilization and load, full-SID collisions, and temporal stability, enabling codebook candidates to be screened before costly downstream model training. Across the two public datasets, the proposed architecture improves mean Recall@10 by 5.0\%--8.8\% and mean NDCG@10 by 4.1\%--5.1\% for OneRec-V1. The corresponding gains for OneRec-V2 are 7.1\%--8.7\% and 3.8\%--8.5\%. Under the fixed-date KuaiRec evaluation, dynamic updating further improves Recall@10 by 1.4\% and 2.7\%, and NDCG@10 by 7.0\% and 2.7\%, for OneRec-V1 and OneRec-V2, respectively. In industrial codebook evaluation, the dynamic codebook improves reconstruction similarity by 1.50\% while changing only 0.4288\% of item-level SID1 assignments in the fixed-date KuaiRec update. Across the decoder, LazyAR, and MTP serving architectures, the shorter SID reduces estimated autoregressive-decoding FLOPs by 47.93\%--48.70\% and raises single-card QPS by 28.57\%--47.0\%. In a five-day online A/B test serving 2.5\% of production traffic, the two-level SID improves the primary consumption metric by 0.792\%. Together, these components provide a practical framework for simplifying, updating, and evaluating SID codebooks in industrial generative recommendation.

\bibliographystyle{plainnat}
\bibliography{references}

\appendix
\section{Inference-Cost Derivation}
\label{app:serving_cost}

We count one multiplication and one addition as two floating-point operations. Thus, multiplying $\mathbf{A}\in\mathbb{R}^{m\times k}$ by $\mathbf{B}\in\mathbb{R}^{k\times n}$ requires approximately
\begin{equation}
  F_{\mathrm{matmul}}(m,k,n)=2mkn
  \label{eq:app_matmul_flops}
\end{equation}
FLOPs. Lower-order element-wise operations are omitted.

Let $L$ denote the number of decoder layers, $d$ the model width, $h$ the feed-forward width, and $N$ the cross-attention context length. At decoding step $j$, let $b_j$ be the number of beams entering the decoder and $V_j$ the output vocabulary size. The cross-attention keys and values are projected once per layer and then cached, giving the shared cost
\begin{equation}
  F_{\mathrm{cross\text{-}KV}}=4LNd^2.
  \label{eq:app_cross_kv_flops}
\end{equation}
For one decoding step, the cross-attention query and output projections cost $4b_jLd^2$, attention against the cached context costs $4b_jLNd$, and the associated SwiGLU feed-forward block costs $6b_jLdh$. With self-attention KV caching, only the current token is projected. Its query--key--value and output projections cost $8b_jLd^2$, attention over the cached prefix costs $4b_jLjd$, and the second feed-forward block costs $6b_jLdh$. The vocabulary projection adds $2b_jdV_j$. Hence, the step cost is
\begin{equation}
  C_j
  =b_jL\left[12d^2+12dh+4d(N+j)\right]
  +2b_jdV_j.
  \label{eq:app_decode_step_flops}
\end{equation}

For an SID containing $q$ generated tokens, the decoder-core cost is
\begin{equation}
  F_q
  =F_{\mathrm{cross\text{-}KV}}
  +\sum_{j=1}^{q}C_j.
  \label{eq:app_q_level_flops}
\end{equation}
After removing the final autoregressive level, it becomes
\begin{equation}
  F_{q-1}
  =F_{\mathrm{cross\text{-}KV}}
  +\sum_{j=1}^{q-1}C'_j,
  \label{eq:app_reduced_level_flops}
\end{equation}
where $C'_j$ uses the beam count and vocabulary size of the shortened SID. The reported decoder-core reduction is $1-F_{q-1}/F_q$. Computation shared by the two SID structures is excluded from both terms.

KV-cache movement is reported separately because memory access is not a floating-point operation. Let $\rho$ denote the bytes per cached element. The cross-attention cache occupies $2\rho LNd$ bytes, and its logical reads across beam search are
\begin{equation}
  T_{\mathrm{cross},q}
  =2\rho LNd\sum_{j=1}^{q}b_j.
  \label{eq:app_cross_kv_traffic}
\end{equation}
The corresponding quantity for the shortened SID replaces $q$ and $b_j$ with its reduced decoding schedule. Self-attention cache reads scale as $2\rho Ld\sum_j b_j\,j$, while cache gathering after beam selection adds implementation-dependent memory traffic. These quantities are logical accesses rather than measured device-memory transactions, since kernel tiling and on-chip reuse are hardware dependent.

\section{OneRec Public-Data Model Configurations}
\label{app:onerec_configs}

Tables~\ref{tab:onerec_v1_configs} and~\ref{tab:onerec_v2_configs} report the exact architecture settings used for the nominal OneRec scales. These are public-data adaptations of the corresponding architectures rather than parameter-exact reproductions of the production systems.

\begin{table}[htbp]
  \centering
  \small
  \caption{OneRec-V1 public-data configurations. All variants use 24 experts and activate two experts per token.}
  \label{tab:onerec_v1_configs}
  \begin{tabular}{@{}lrrrrrr@{}}
    \toprule
    Scale & $d_{\mathrm{model}}$ & Enc. layers & Dec. layers & Heads & Enc. FFN & Expert FFN \\
    \midrule
    0.1B & 512 & 6 & 6 & 8 & 2,048 & 256 \\
    0.2B & 640 & 8 & 8 & 10 & 2,560 & 320 \\
    0.3B & 768 & 8 & 10 & 12 & 3,072 & 320 \\
    0.4B & 896 & 8 & 8 & 14 & 3,584 & 448 \\
    0.5B & 1,024 & 8 & 8 & 16 & 4,096 & 512 \\
    0.6B & 1,024 & 10 & 10 & 16 & 4,096 & 512 \\
    0.7B & 1,152 & 9 & 9 & 18 & 4,608 & 576 \\
    \bottomrule
  \end{tabular}
\end{table}

\begin{table}[htbp]
  \centering
  \small
  \caption{OneRec-V2 public-data configurations.}
  \label{tab:onerec_v2_configs}
  \begin{tabular}{@{}lrrrrr@{}}
    \toprule
    Scale & $d_{\mathrm{model}}$ & Layers & Attention heads & KV heads & FFN \\
    \midrule
    0.1B & 768 & 12 & 12 & 4 & 2,048 \\
    0.2B & 1,024 & 16 & 16 & 4 & 2,048 \\
    0.3B & 1,024 & 20 & 16 & 4 & 3,072 \\
    0.4B & 1,280 & 18 & 20 & 4 & 3,072 \\
    0.5B & 1,280 & 20 & 20 & 4 & 4,096 \\
    0.6B & 1,280 & 24 & 20 & 4 & 4,096 \\
    0.7B & 1,536 & 22 & 24 & 4 & 4,096 \\
    \bottomrule
  \end{tabular}
\end{table}

\section{OneRec Results on Amazon Beauty}
\label{app:onerec_amazon_results}

Tables~\ref{tab:amazon_onerec_v1_results} and~\ref{tab:amazon_onerec_v2_results} provide the exact values underlying Figure~\ref{fig:amazon_onerec_scaling}. Each entry is the mean $\pm$ sample standard deviation over seeds $\{7,42,2024\}$.

\begin{table}[htbp]
  \centering
  \small
  \caption{OneRec-V1 results on Amazon Reviews 2014 (Beauty).}
  \label{tab:amazon_onerec_v1_results}
  \resizebox{\linewidth}{!}{%
  \begin{tabular}{@{}llcccc@{}}
    \toprule
    Scale & SID & Recall@5 & Recall@10 & NDCG@5 & NDCG@10 \\
    \midrule
    0.1B & S3 & $0.0243\pm0.0004$ & $0.0411\pm0.0016$ & $0.0152\pm0.0002$ & $0.0206\pm0.0006$ \\
    0.1B & S2 & $0.0253\pm0.0005$ & $0.0429\pm0.0006$ & $0.0156\pm0.0003$ & $0.0212\pm0.0004$ \\
    \addlinespace
    0.2B & S3 & $0.0241\pm0.0008$ & $0.0397\pm0.0011$ & $0.0150\pm0.0006$ & $0.0200\pm0.0007$ \\
    0.2B & S2 & $0.0252\pm0.0017$ & $0.0415\pm0.0013$ & $0.0155\pm0.0011$ & $0.0208\pm0.0010$ \\
    \addlinespace
    0.3B & S3 & $0.0219\pm0.0014$ & $0.0372\pm0.0008$ & $0.0138\pm0.0010$ & $0.0187\pm0.0009$ \\
    0.3B & S2 & $0.0241\pm0.0002$ & $0.0402\pm0.0015$ & $0.0148\pm0.0001$ & $0.0199\pm0.0007$ \\
    \addlinespace
    0.4B & S3 & $0.0236\pm0.0002$ & $0.0388\pm0.0010$ & $0.0143\pm0.0004$ & $0.0192\pm0.0004$ \\
    0.4B & S2 & $0.0235\pm0.0001$ & $0.0395\pm0.0008$ & $0.0146\pm0.0001$ & $0.0197\pm0.0001$ \\
    \addlinespace
    0.5B & S3 & $0.0220\pm0.0017$ & $0.0366\pm0.0017$ & $0.0136\pm0.0008$ & $0.0183\pm0.0008$ \\
    0.5B & S2 & $0.0232\pm0.0013$ & $0.0399\pm0.0014$ & $0.0145\pm0.0010$ & $0.0199\pm0.0011$ \\
    \addlinespace
    0.6B & S3 & $0.0226\pm0.0009$ & $0.0369\pm0.0006$ & $0.0141\pm0.0006$ & $0.0187\pm0.0005$ \\
    0.6B & S2 & $0.0237\pm0.0007$ & $0.0383\pm0.0011$ & $0.0146\pm0.0006$ & $0.0193\pm0.0007$ \\
    \addlinespace
    0.7B & S3 & $0.0222\pm0.0007$ & $0.0371\pm0.0009$ & $0.0137\pm0.0004$ & $0.0185\pm0.0002$ \\
    0.7B & S2 & $0.0223\pm0.0018$ & $0.0383\pm0.0027$ & $0.0136\pm0.0009$ & $0.0187\pm0.0011$ \\
    \bottomrule
  \end{tabular}}
\end{table}

\begin{table}[htbp]
  \centering
  \small
  \caption{OneRec-V2 results on Amazon Reviews 2014 (Beauty).}
  \label{tab:amazon_onerec_v2_results}
  \resizebox{\linewidth}{!}{%
  \begin{tabular}{@{}llcccc@{}}
    \toprule
    Scale & SID & Recall@5 & Recall@10 & NDCG@5 & NDCG@10 \\
    \midrule
    0.1B & S3 & $0.0308\pm0.0013$ & $0.0476\pm0.0022$ & $0.0196\pm0.0006$ & $0.0250\pm0.0008$ \\
    0.1B & S2 & $0.0315\pm0.0014$ & $0.0509\pm0.0008$ & $0.0203\pm0.0010$ & $0.0266\pm0.0008$ \\
    \addlinespace
    0.2B & S3 & $0.0280\pm0.0003$ & $0.0444\pm0.0014$ & $0.0178\pm0.0004$ & $0.0231\pm0.0004$ \\
    0.2B & S2 & $0.0301\pm0.0014$ & $0.0492\pm0.0009$ & $0.0193\pm0.0010$ & $0.0255\pm0.0008$ \\
    \addlinespace
    0.3B & S3 & $0.0272\pm0.0007$ & $0.0437\pm0.0014$ & $0.0175\pm0.0003$ & $0.0228\pm0.0006$ \\
    0.3B & S2 & $0.0316\pm0.0008$ & $0.0503\pm0.0017$ & $0.0202\pm0.0004$ & $0.0263\pm0.0007$ \\
    \addlinespace
    0.4B & S3 & $0.0280\pm0.0009$ & $0.0446\pm0.0010$ & $0.0180\pm0.0007$ & $0.0233\pm0.0004$ \\
    0.4B & S2 & $0.0310\pm0.0008$ & $0.0488\pm0.0009$ & $0.0200\pm0.0003$ & $0.0257\pm0.0003$ \\
    \addlinespace
    0.5B & S3 & $0.0268\pm0.0010$ & $0.0430\pm0.0009$ & $0.0173\pm0.0005$ & $0.0225\pm0.0005$ \\
    0.5B & S2 & $0.0299\pm0.0016$ & $0.0477\pm0.0025$ & $0.0192\pm0.0012$ & $0.0249\pm0.0013$ \\
    \addlinespace
    0.6B & S3 & $0.0287\pm0.0030$ & $0.0448\pm0.0027$ & $0.0185\pm0.0019$ & $0.0237\pm0.0018$ \\
    0.6B & S2 & $0.0296\pm0.0015$ & $0.0471\pm0.0021$ & $0.0189\pm0.0011$ & $0.0245\pm0.0015$ \\
    \addlinespace
    0.7B & S3 & $0.0278\pm0.0017$ & $0.0440\pm0.0025$ & $0.0177\pm0.0012$ & $0.0230\pm0.0015$ \\
    0.7B & S2 & $0.0277\pm0.0012$ & $0.0450\pm0.0022$ & $0.0181\pm0.0007$ & $0.0237\pm0.0009$ \\
    \bottomrule
  \end{tabular}}
\end{table}

\section{OneRec Results on KuaiRec}
\label{app:onerec_kuairec_results}

Tables~\ref{tab:kuairec_onerec_v1_results} and~\ref{tab:kuairec_onerec_v2_results} provide the exact values underlying Figure~\ref{fig:kuairec_onerec_scaling}. Each entry is the mean $\pm$ sample standard deviation over seeds $\{7,42,2024\}$.

\newcommand{\kuairecVOneCaption}{OneRec-V1 results on KuaiRec.}
\newcommand{\kuairecVTwoCaption}{OneRec-V2 results on KuaiRec.}
\begin{table}[htbp]
  \centering
  \scriptsize
  \caption{\kuairecVOneCaption}
  \label{tab:kuairec_onerec_v1_results}
  \resizebox{\linewidth}{!}{%
  \begin{tabular}{@{}llcccc@{}}
    \toprule
    Scale & SID & Recall@5 & Recall@10 & NDCG@5 & NDCG@10 \\
    \midrule
    0.1B & S3 & $0.0826\pm0.0025$ & $0.1098\pm0.0037$ & $0.0539\pm0.0036$ & $0.0628\pm0.0040$ \\
    0.1B & S2 & $0.0854\pm0.0037$ & $0.1196\pm0.0025$ & $0.0553\pm0.0021$ & $0.0664\pm0.0017$ \\
    0.1B & PV-S2 & $0.0846\pm0.0047$ & $0.1179\pm0.0057$ & $0.0556\pm0.0030$ & $0.0664\pm0.0033$ \\
    \addlinespace
    0.2B & S3 & $0.0733\pm0.0057$ & $0.1001\pm0.0062$ & $0.0483\pm0.0029$ & $0.0569\pm0.0031$ \\
    0.2B & S2 & $0.0735\pm0.0122$ & $0.1053\pm0.0062$ & $0.0456\pm0.0065$ & $0.0559\pm0.0045$ \\
    0.2B & PV-S2 & $0.0719\pm0.0033$ & $0.1039\pm0.0033$ & $0.0466\pm0.0021$ & $0.0570\pm0.0021$ \\
    \addlinespace
    0.3B & S3 & $0.0625\pm0.0026$ & $0.0909\pm0.0037$ & $0.0397\pm0.0024$ & $0.0490\pm0.0027$ \\
    0.3B & S2 & $0.0728\pm0.0108$ & $0.1040\pm0.0100$ & $0.0463\pm0.0072$ & $0.0564\pm0.0069$ \\
    0.3B & PV-S2 & $0.0754\pm0.0072$ & $0.1051\pm0.0068$ & $0.0470\pm0.0057$ & $0.0567\pm0.0056$ \\
    \addlinespace
    0.4B & S3 & $0.0687\pm0.0063$ & $0.0942\pm0.0053$ & $0.0429\pm0.0047$ & $0.0512\pm0.0044$ \\
    0.4B & S2 & $0.0704\pm0.0076$ & $0.1055\pm0.0071$ & $0.0433\pm0.0042$ & $0.0548\pm0.0040$ \\
    0.4B & PV-S2 & $0.0746\pm0.0043$ & $0.1056\pm0.0038$ & $0.0465\pm0.0027$ & $0.0566\pm0.0025$ \\
    \addlinespace
    0.5B & S3 & $0.0654\pm0.0116$ & $0.0962\pm0.0116$ & $0.0421\pm0.0091$ & $0.0521\pm0.0091$ \\
    0.5B & S2 & $0.0696\pm0.0047$ & $0.1027\pm0.0070$ & $0.0442\pm0.0030$ & $0.0549\pm0.0039$ \\
    0.5B & PV-S2 & $0.0687\pm0.0017$ & $0.1009\pm0.0024$ & $0.0439\pm0.0005$ & $0.0543\pm0.0005$ \\
    \addlinespace
    0.6B & S3 & $0.0662\pm0.0057$ & $0.0916\pm0.0037$ & $0.0422\pm0.0045$ & $0.0505\pm0.0037$ \\
    0.6B & S2 & $0.0675\pm0.0044$ & $0.1024\pm0.0057$ & $0.0418\pm0.0031$ & $0.0531\pm0.0034$ \\
    0.6B & PV-S2 & $0.0668\pm0.0033$ & $0.0997\pm0.0021$ & $0.0415\pm0.0022$ & $0.0521\pm0.0017$ \\
    \addlinespace
    0.7B & S3 & $0.0652\pm0.0029$ & $0.0926\pm0.0065$ & $0.0412\pm0.0017$ & $0.0502\pm0.0030$ \\
    0.7B & S2 & $0.0641\pm0.0047$ & $0.0948\pm0.0046$ & $0.0404\pm0.0032$ & $0.0503\pm0.0031$ \\
    0.7B & PV-S2 & $0.0593\pm0.0012$ & $0.0889\pm0.0009$ & $0.0380\pm0.0026$ & $0.0476\pm0.0025$ \\
    \bottomrule
  \end{tabular}}
\end{table}

\begin{table}[htbp]
  \centering
  \scriptsize
  \caption{\kuairecVTwoCaption}
  \label{tab:kuairec_onerec_v2_results}
  \resizebox{\linewidth}{!}{%
  \begin{tabular}{@{}llcccc@{}}
    \toprule
    Scale & SID & Recall@5 & Recall@10 & NDCG@5 & NDCG@10 \\
    \midrule
    0.1B & S3 & $0.0931\pm0.0024$ & $0.1244\pm0.0020$ & $0.0656\pm0.0013$ & $0.0757\pm0.0013$ \\
    0.1B & S2 & $0.1014\pm0.0067$ & $0.1351\pm0.0058$ & $0.0713\pm0.0049$ & $0.0823\pm0.0046$ \\
    0.1B & PV-S2 & $0.0977\pm0.0055$ & $0.1335\pm0.0058$ & $0.0692\pm0.0038$ & $0.0808\pm0.0040$ \\
    \addlinespace
    0.2B & S3 & $0.0964\pm0.0036$ & $0.1271\pm0.0025$ & $0.0663\pm0.0037$ & $0.0762\pm0.0029$ \\
    0.2B & S2 & $0.0968\pm0.0069$ & $0.1342\pm0.0056$ & $0.0659\pm0.0020$ & $0.0781\pm0.0017$ \\
    0.2B & PV-S2 & $0.0966\pm0.0036$ & $0.1322\pm0.0020$ & $0.0676\pm0.0032$ & $0.0791\pm0.0027$ \\
    \addlinespace
    0.3B & S3 & $0.0935\pm0.0068$ & $0.1234\pm0.0036$ & $0.0654\pm0.0060$ & $0.0751\pm0.0049$ \\
    0.3B & S2 & $0.0956\pm0.0019$ & $0.1307\pm0.0019$ & $0.0647\pm0.0006$ & $0.0761\pm0.0008$ \\
    0.3B & PV-S2 & $0.0949\pm0.0055$ & $0.1277\pm0.0063$ & $0.0654\pm0.0029$ & $0.0761\pm0.0032$ \\
    \addlinespace
    0.4B & S3 & $0.0896\pm0.0047$ & $0.1169\pm0.0015$ & $0.0610\pm0.0048$ & $0.0699\pm0.0037$ \\
    0.4B & S2 & $0.0955\pm0.0030$ & $0.1322\pm0.0010$ & $0.0652\pm0.0026$ & $0.0771\pm0.0014$ \\
    0.4B & PV-S2 & $0.0924\pm0.0085$ & $0.1272\pm0.0039$ & $0.0637\pm0.0047$ & $0.0750\pm0.0032$ \\
    \addlinespace
    0.5B & S3 & $0.0893\pm0.0070$ & $0.1197\pm0.0053$ & $0.0612\pm0.0049$ & $0.0711\pm0.0044$ \\
    0.5B & S2 & $0.0985\pm0.0054$ & $0.1338\pm0.0049$ & $0.0664\pm0.0057$ & $0.0779\pm0.0055$ \\
    0.5B & PV-S2 & $0.0950\pm0.0056$ & $0.1289\pm0.0028$ & $0.0654\pm0.0046$ & $0.0764\pm0.0035$ \\
    \addlinespace
    0.6B & S3 & $0.0965\pm0.0010$ & $0.1250\pm0.0042$ & $0.0692\pm0.0015$ & $0.0784\pm0.0002$ \\
    0.6B & S2 & $0.0955\pm0.0047$ & $0.1269\pm0.0059$ & $0.0651\pm0.0034$ & $0.0752\pm0.0038$ \\
    0.6B & PV-S2 & $0.0975\pm0.0034$ & $0.1370\pm0.0027$ & $0.0674\pm0.0036$ & $0.0802\pm0.0027$ \\
    \addlinespace
    0.7B & S3 & $0.0916\pm0.0039$ & $0.1206\pm0.0073$ & $0.0641\pm0.0040$ & $0.0735\pm0.0048$ \\
    0.7B & S2 & $0.0897\pm0.0030$ & $0.1249\pm0.0010$ & $0.0614\pm0.0024$ & $0.0729\pm0.0019$ \\
    0.7B & PV-S2 & $0.0960\pm0.0027$ & $0.1299\pm0.0037$ & $0.0662\pm0.0037$ & $0.0771\pm0.0039$ \\
    \bottomrule
  \end{tabular}}
\end{table}

\section{OneRec Fixed-Date Static--Dynamic Results on KuaiRec}
\label{app:onerec_kuairec_temporal_results}

Tables~\ref{tab:kuairec_temporal_onerec_v1_results} and~\ref{tab:kuairec_temporal_onerec_v2_results} provide the exact values underlying Figure~\ref{fig:kuairec_onerec_temporal_scaling}. Each entry is the mean $\pm$ sample standard deviation over seeds $\{7,42,2024\}$.

\begin{table}[htbp]
  \centering
  \small
  \caption{OneRec-V1 results in the fixed-date KuaiRec comparison.}
  \label{tab:kuairec_temporal_onerec_v1_results}
  \resizebox{\linewidth}{!}{%
  \begin{tabular}{@{}llcccc@{}}
    \toprule
    Scale & Codebook & Recall@5 & Recall@10 & NDCG@5 & NDCG@10 \\
    \midrule
    0.1B & Static C1 & $0.1268\pm0.0076$ & $0.1467\pm0.0082$ & $0.0972\pm0.0089$ & $0.1046\pm0.0106$ \\
    0.1B & Dynamic C2 & $0.1328\pm0.0104$ & $0.1567\pm0.0084$ & $0.1028\pm0.0056$ & $0.1115\pm0.0044$ \\
    \addlinespace
    0.2B & Static C1 & $0.1138\pm0.0170$ & $0.1388\pm0.0118$ & $0.0832\pm0.0208$ & $0.0913\pm0.0189$ \\
    0.2B & Dynamic C2 & $0.1102\pm0.0206$ & $0.1374\pm0.0197$ & $0.0832\pm0.0197$ & $0.0921\pm0.0194$ \\
    \addlinespace
    0.3B & Static C1 & $0.1198\pm0.0069$ & $0.1445\pm0.0053$ & $0.0875\pm0.0078$ & $0.0956\pm0.0074$ \\
    0.3B & Dynamic C2 & $0.1212\pm0.0111$ & $0.1515\pm0.0128$ & $0.0959\pm0.0126$ & $0.1061\pm0.0132$ \\
    \addlinespace
    0.4B & Static C1 & $0.0994\pm0.0075$ & $0.1300\pm0.0097$ & $0.0707\pm0.0065$ & $0.0807\pm0.0073$ \\
    0.4B & Dynamic C2 & $0.1003\pm0.0124$ & $0.1264\pm0.0137$ & $0.0791\pm0.0029$ & $0.0875\pm0.0029$ \\
    \addlinespace
    0.5B & Static C1 & $0.1088\pm0.0126$ & $0.1377\pm0.0182$ & $0.0685\pm0.0196$ & $0.0790\pm0.0200$ \\
    0.5B & Dynamic C2 & $0.1049\pm0.0077$ & $0.1359\pm0.0125$ & $0.0721\pm0.0047$ & $0.0822\pm0.0063$ \\
    \addlinespace
    0.6B & Static C1 & $0.1024\pm0.0202$ & $0.1317\pm0.0130$ & $0.0621\pm0.0090$ & $0.0717\pm0.0064$ \\
    0.6B & Dynamic C2 & $0.1092\pm0.0186$ & $0.1355\pm0.0135$ & $0.0746\pm0.0038$ & $0.0837\pm0.0035$ \\
    \addlinespace
    0.7B & Static C1 & $0.0918\pm0.0156$ & $0.1199\pm0.0137$ & $0.0539\pm0.0161$ & $0.0641\pm0.0144$ \\
    0.7B & Dynamic C2 & $0.0937\pm0.0173$ & $0.1194\pm0.0220$ & $0.0578\pm0.0130$ & $0.0653\pm0.0143$ \\
    \bottomrule
  \end{tabular}}
\end{table}

\begin{table}[htbp]
  \centering
  \small
  \caption{OneRec-V2 results in the fixed-date KuaiRec comparison.}
  \label{tab:kuairec_temporal_onerec_v2_results}
  \resizebox{\linewidth}{!}{%
  \begin{tabular}{@{}llcccc@{}}
    \toprule
    Scale & Codebook & Recall@5 & Recall@10 & NDCG@5 & NDCG@10 \\
    \midrule
    0.1B & Static C1 & $0.1406\pm0.0284$ & $0.1718\pm0.0334$ & $0.1095\pm0.0235$ & $0.1197\pm0.0249$ \\
    0.1B & Dynamic C2 & $0.1432\pm0.0066$ & $0.1804\pm0.0066$ & $0.1078\pm0.0080$ & $0.1200\pm0.0077$ \\
    \addlinespace
    0.2B & Static C1 & $0.1366\pm0.0131$ & $0.1659\pm0.0099$ & $0.1090\pm0.0131$ & $0.1186\pm0.0122$ \\
    0.2B & Dynamic C2 & $0.1386\pm0.0176$ & $0.1728\pm0.0140$ & $0.1067\pm0.0160$ & $0.1179\pm0.0149$ \\
    \addlinespace
    0.3B & Static C1 & $0.1391\pm0.0061$ & $0.1684\pm0.0060$ & $0.1072\pm0.0123$ & $0.1168\pm0.0124$ \\
    0.3B & Dynamic C2 & $0.1465\pm0.0074$ & $0.1838\pm0.0140$ & $0.1137\pm0.0026$ & $0.1259\pm0.0050$ \\
    \addlinespace
    0.4B & Static C1 & $0.1404\pm0.0148$ & $0.1762\pm0.0160$ & $0.1010\pm0.0147$ & $0.1127\pm0.0148$ \\
    0.4B & Dynamic C2 & $0.1424\pm0.0276$ & $0.1680\pm0.0401$ & $0.1149\pm0.0265$ & $0.1254\pm0.0269$ \\
    \addlinespace
    0.5B & Static C1 & $0.1260\pm0.0141$ & $0.1638\pm0.0214$ & $0.0928\pm0.0079$ & $0.1051\pm0.0102$ \\
    0.5B & Dynamic C2 & $0.1272\pm0.0042$ & $0.1610\pm0.0037$ & $0.0921\pm0.0022$ & $0.1042\pm0.0032$ \\
    \addlinespace
    0.6B & Static C1 & $0.1248\pm0.0085$ & $0.1546\pm0.0124$ & $0.0934\pm0.0070$ & $0.1032\pm0.0083$ \\
    0.6B & Dynamic C2 & $0.1217\pm0.0074$ & $0.1508\pm0.0065$ & $0.0886\pm0.0062$ & $0.0992\pm0.0071$ \\
    \addlinespace
    0.7B & Static C1 & $0.1299\pm0.0078$ & $0.1605\pm0.0114$ & $0.0981\pm0.0059$ & $0.1073\pm0.0052$ \\
    0.7B & Dynamic C2 & $0.1364\pm0.0065$ & $0.1759\pm0.0130$ & $0.0970\pm0.0087$ & $0.1117\pm0.0030$ \\
    \bottomrule
  \end{tabular}}
\end{table}

\end{document}